\documentclass[showpacs,twocolumn,superscriptaddress,aps]{revtex4}%
\usepackage{amsmath}
\usepackage{graphicx}
\usepackage{amsfonts}
\usepackage{amssymb}%
\begin{document}
\title{Weaving independently generated photons into an arbitrary graph state}
\author{Qing Lin}
\email{qlin@mail.ustc.edu.cn}
\affiliation{College of Information Science and Engineering, Huaqiao University (Xiamen),
Xiamen 361021, China}
\author{Bing He}
\email{bhe4@ucmerced.edu}
\affiliation{University of California, Merced, 5200 North Lake Road, Merced, CA 95343, USA}
\affiliation{Institute for Quantum Information Science, University of Calgary, Alberta T2N
1N4, Canada}

\pacs{03.67.Lx, 42.50.Ex}

\begin{abstract}
The controlled Z (CZ) operations acting separately on pairs of qubits are
commonly adopted in the schemes of generating graph states, the multi-partite
entangled states for the one-way quantum computing. For this purpose, we propose
a setup of cascade CZ operation on a whole group of qubits in sequence.
The operation of the setup starts with entangling an ancilla photon to the
first photon as qubit, and this ancilla automatically moves from one
entanglement link to another in assisting the formation of a string in graph
states. The generation of some special types of graph states, such as the
three-dimensional ones, can be greatly simplified in this approach. The setup presented
uses weak nonlinearities, but an implementation using probabilistic linear optics is also possible.

\end{abstract}
\maketitle

\section{Introduction}

The one-way quantum computing \cite{cluster1,cluster2,cluster3} attracts wide
attention for its efficiency and simplicity. Different from the traditional
circuit-based quantum computing, it only works with single-qubit measurements
on the apriori prepared multi-entangled states called graph states or cluster
states. The graph states form an important class of multi-partite entangled
states. In addition to the two-dimensional (2D) graph states, three
dimensional (3D) graph states have been suggested for fault-tolerant one-way
quantum computing \cite{3D}. How to efficiently generate these graph states is
the main problem in realizing practical one-way quantum computing.

Photons have long coherence time and interact weakly with their environment.
Photonic qubits, including the discrete ones (see e.g. \cite{Nielsen1, Browne,
Gilbert, dc-c1}) and the continuous variable (CV) ones (see e.g. \cite{cv-1,
cv-2, cv-3}), are among the first candidates for the one-way quantum
computing. Although there have existed numerous proof of principle experiments
\cite{experiments1, experiments2, k, pan1, y, pan2} to demonstrate the
generation of graph states of a few photons, these linear optical approaches
are not efficient enough for making graph states of many qubits due to their
limited success probabilities in basic entangling operations. Deterministic
gates employing photonic nonlinearity are necessary for the practical one-way
quantum computing. The straightforward way of entangling photonic qubits is to
apply deterministic controlled-Z (CZ) gate working with strong photon-photon
interaction in nonlinear media. However, in addition to the generic weak
interaction between photons and the accompanying decoherence from losses in
media, the realization of high-quality photon-photon gates is hindered by
physical limitations such as multi-mode effects \cite{shap1, bana, multi,
qft}. An alternative for realizing deterministic photonic gates is the weak
nonlinearity between photons and coherent states with large amplitude
\cite{Nemoto, Barrett}. So far the schemes based on weak nonlinearity have
been proposed for generating 2D graph states of atomic qubits \cite{Spiller1,
Spiller2, hors} or photonic qubits \cite{Spiller3, Lin3}. The operations in
generating graph states of large number of qubits should be also optimized, so
that the demanded resources could be as few as possible \cite{hors}. This
problem is more relevant to graph states of photonic qubits, which should be
prepared quickly if given no perfect quantum memory.

To have a clearer picture of the problem, we look at how a graph state is
built up from the inputs. Most of theoretical works adopt the link by link
entanglement connection between qubits by separate CZ operations. Given weak
nonlinearity, each CZ operation requires two entangler or parity gate
operations, which consume an ancilla photon, respectively \cite{Nemoto}.
Another approach for speeding up graph state generation is to manufacture a
target graph state with the building blocks initially prepared from the
elementary qubits. For example, in \cite{Lin3} we propose a procedure of
generating 2D graph states in this fashion; the box-shaped building blocks are
first prepared from qubits with entangler operations together with CZ
operations, and then the building blocks are assembled to the target graph
state by CZ operations. This approach reduces most of CZ operations to
deterministic entangler or parity gate operations. To reduce the total
preparation time, however, a considerable number of entanglers should be used
to prepare the building blocks simultaneously. This necessitates a trade-off
of the preparation times for the preparation resources.

Here we present an architecture of \textit{cascade CZ operation} to solve the
problem. By cascade CZ operation we mean that the CZ operations to entangle
qubits into a string are bundled together and performed by a single setup
called cascade entangler. Such operation is assisted by only one ancilla
photon called \textit{spider photon}, which moves from one entanglement link
to another throughout a cascade CZ operation and acts like a spider weaving
qubits into a graph state. With cascade CZ operations, the number of
facilities for simultaneous operations in generating a graph state, as well as
the corresponding ancilla photon number, could be reduced to that of connected
strings in the graph state. Moreover, because of the flexible passage of a
spider photon, such setup is able to generate an arbitrary graph state.

The rest of the paper is organized as follows. In Sec. II we provide a
detailed description of how cascade entangler works for our purpose. The
generation of clusters states is illustrated in Sec. III with examples, where
we emphasize the advantage of the approach generating 3D graph states.
Finally, in Sec. IV we give more discussion on the core element, cross-Kerr
nonlinearity, in our setup, and conclude the paper.

\section{Cascade entangler and cascade CZ operation}

The design of cascade entangler is based on the following decomposition of a
general graph state \cite{review}:
\begin{align}
|G\rangle &  =\prod_{(i,j)\in E}CZ_{i,j}|+\rangle^{\otimes V}\nonumber\\
&  =\prod_{(i,j)\in E_{1}}CZ_{i,j}|+\rangle^{\otimes V_{1}}\cdots
\prod_{(i,j)\in E_{n}}CZ_{i,j}|+\rangle^{\otimes V_{n}}, \label{general}%
\end{align}
where $|\pm\rangle=\frac{1}{\sqrt{2}}(|0\rangle\pm|1\rangle)$ is a qubit on a
vertex in the sets $V_{i}$, and $CZ_{i,j}$ denotes the CZ operation over the
link between vertex $i$ and $j$. By the above expression, a general graph
state is decomposed into a product of the connected string structures $E_{i}$.
Note that the CZ operations in the above equation commute and the
decomposition to the products of the different connected strings is not
unique. The setup is designed to successively entangle the qubits to the
string structures $E_{i}$, and the creation of these strings is assisted by a
spider photon. If a graph state is in the shape of a single string $E$ (the
number of set $E_{i}$ in Eq. (\ref{general}) equals to one), the setup will
directly generate the graph state by the repeated entangling operations on the
input qubits. Given $n$ simultaneous operations to generate the strings
$E_{i}$, a graph state could be created in an efficiently way.

We illustrate the operation of a cascade entangler with the following input
state involving photons $p$, $r$ and the spider photon $a$ as the ancilla (the
qubits are encoded with the polarization modes, $|0\rangle\equiv|H\rangle$,
$|1\rangle\equiv|V\rangle$):
\begin{align}
|\Psi\rangle_{in}  &  = (\left\vert \psi_{1}\right\rangle \left\vert
0\right\rangle _{p} \left\vert +\right\rangle _{a} +\left\vert \psi
_{2}\right\rangle \left\vert 1\right\rangle _{p} \left\vert -\right\rangle
_{a} )\otimes(|0\rangle_{r}+|1\rangle_{r})\nonumber\\
&  = \big( \left\vert \psi_{1}\right\rangle \left\vert 00\right\rangle
\left\vert +\right\rangle +\left\vert \psi_{1}\right\rangle \left\vert
01\right\rangle \left\vert +\right\rangle +\left\vert \psi_{2}\right\rangle
\left\vert 10\right\rangle \left\vert -\right\rangle \nonumber\\
&  +\left\vert \psi_{2}\right\rangle \left\vert 11\right\rangle \left\vert
-\right\rangle \big ) _{pra}, \label{in}%
\end{align}
where $\left\vert \psi_{i}\right\rangle $, for $i=1,2$, are the proper
unnormalized pure states involving other photons than $p$. Here we assume that
the spider photon $a$ has been entangled to the finished piece of a graph
state, and photon $r$ will be attached to this piece. The already prepared
piece $\left\vert \psi_{1}\right\rangle \left\vert 0\right\rangle
_{p}+\left\vert \psi_{2}\right\rangle \left\vert 1\right\rangle _{p}$ can be a
general graph state, but the spider photon $a$ is always entangled to the
neighboring photon in the finished piece (photon $p$, for example) in a
specific way, i.e., $|+\rangle_{a}$ ($|-\rangle_{a}$) is guaranteed to be in
the same terms with $|0\rangle_{p}$ ($|1\rangle_{p}$); see Eq. (\ref{in}). The
purpose for such choice in cascade CZ operation will be seen below.

\begin{figure}[ptb]
\includegraphics[width=7.0cm]{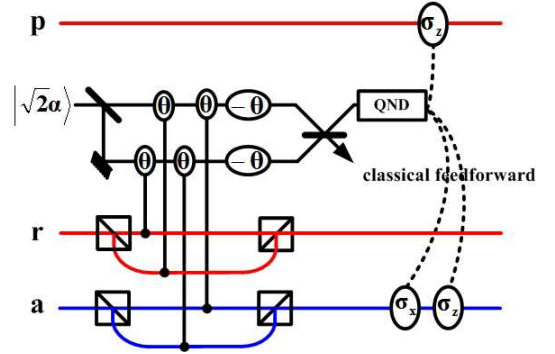}\caption{(color online) Cascade
entangler based on cross-phase modulation between photons and coherent states.
The two qubus beams are coupled to the $\left\vert 0\right\rangle $,
$\left\vert 1\right\rangle $ of the photons $r$ and $a$ as indicated. The
ancilla photon (spider photon) is denoted as $a$. A phase shift $-\theta$ is
applied to the qubus beams respectively. The qubus beam is divided and merged
by 50:50 beam splitters. The single-qubit operations $\sigma_{x}$ and
$\sigma_{z}$ are conditionally implemented according to the detection results
of the QND module.}%
\end{figure}

We begin with using two polarization beam splitters (PBS) to divide the
photons $r$ and $a$ into two spatial modes, respectively. Then two quantum bus
(qubus) beams in the coherent state $\left\vert \alpha\right\rangle $ are
coupled with single photon $r$ and $a$ through the weak cross-Kerr
nonlinearities; see Fig. 1 for the coupling pattern. All induced phase shifts
from cross-phase modulation (XPM) between coherent and single photon state are
assumed to be $\theta$. After that, two phase shifters of $-\theta$ are
respectively applied to the qubus beams to obtain the state
\begin{align}
& ~~~(\left\vert \psi_{1}\right\rangle \left\vert 0\right\rangle
_{p}\left\vert 0\right\rangle _{r}\left\vert 0\right\rangle _{a}+\left\vert
\psi_{1}\right\rangle \left\vert 0\right\rangle _{p}\left\vert 1\right\rangle
_{r}\left\vert 1\right\rangle _{a})\left\vert \alpha\right\rangle \left\vert
\alpha\right\rangle \nonumber\\
&  +(\left\vert \psi_{2}\right\rangle \left\vert 1\right\rangle _{p}\left\vert
0\right\rangle _{r}\left\vert 0\right\rangle _{a}-\left\vert \psi
_{2}\right\rangle \left\vert 1\right\rangle _{p}\left\vert 1\right\rangle
_{r}\left\vert 1\right\rangle _{a})\left\vert \alpha\right\rangle \left\vert
\alpha\right\rangle \nonumber\\
&  +(\left\vert \psi_{1}\right\rangle \left\vert 0\right\rangle _{p}%
-\left\vert \psi_{2}\right\rangle \left\vert 1\right\rangle _{p})\left\vert
0\right\rangle _{r}\left\vert 1\right\rangle _{a}\left\vert \alpha
e^{-i\theta}\right\rangle \left\vert \alpha e^{i\theta}\right\rangle
\nonumber\\
&  +(\left\vert \psi_{1}\right\rangle \left\vert 0\right\rangle _{p}%
+\left\vert \psi_{2}\right\rangle \left\vert 1\right\rangle _{p})\left\vert
1\right\rangle _{r}\left\vert 0\right\rangle _{a}\left\vert \alpha e^{i\theta
}\right\rangle \left\vert \alpha e^{-i\theta}\right\rangle .
\end{align}
Next, one 50:50 beam splitter (BS) performs the transformation $\left\vert
\alpha_{1}\right\rangle \left\vert \alpha_{2}\right\rangle \rightarrow
\left\vert \frac{\alpha_{1}-\alpha_{2}}{\sqrt{2}}\right\rangle \left\vert
\frac{\alpha_{1}+\alpha_{2}}{\sqrt{2}}\right\rangle $ of the qubus coherent states.

A proper output state can be obtained by a continued projection $\left\vert
n\right\rangle \left\langle n\right\vert $ on the qubus beam in the state
$|\pm\sqrt{2}\alpha\sin\theta\rangle$ or $|0\rangle$. If $n=0$, we will have
\begin{align}
&  ~~\left(  \left\vert \psi_{1}\right\rangle \left\vert 0\right\rangle
\left\vert 0\right\rangle \left\vert 0\right\rangle +\left\vert \psi
_{1}\right\rangle \left\vert 0\right\rangle \left\vert 1\right\rangle
\left\vert 1\right\rangle +\left\vert \psi_{2}\right\rangle \left\vert
1\right\rangle \left\vert 0\right\rangle \left\vert 0\right\rangle \right.
\nonumber\\
&  \left.  -\left\vert \psi_{2}\right\rangle \left\vert 1\right\rangle
\left\vert 1\right\rangle \left\vert 1\right\rangle \right)  _{pra};
\end{align}
if $n\neq0$, we can also get the above state by a $\sigma_{z}$ operation on
photon $p$, a $\sigma_{x}$ and $\sigma_{z}$ operation on photon $a$, which are
performed conditionally on the the classically feed-forwarded measurement results.

The projection $\left\vert n\right\rangle \left\langle n\right\vert $ can be
performed by a QND module employing coherent state comparison \cite{module1,
module2}. While the stronger beam of the qubus in the state $|\sqrt{2}%
\alpha\cos\theta\rangle$ or $|\sqrt{2}\alpha\rangle$ will be recycled for the
next entangling operation, the other beam in the state $|\pm\sqrt{2}\alpha
\sin\theta\rangle$ or $|0\rangle$ will be coupled in the module to one of the
beams in the coherent state $|\gamma\rangle$ by a same weak cross-Kerr
nonlinearity, so that the output of the QND module will be obtained from the
process
\begin{align}
&  |\pm\sqrt{2}\alpha\sin\theta\rangle|\gamma\rangle|\gamma\rangle\nonumber\\
&  \rightarrow\sum_{n=0}^{\infty}e^{-|\alpha\sin\theta|^{2}} \frac{(\pm
\sqrt{2}\alpha\sin\theta)^{n}}{\sqrt{n!}}|n\rangle|\frac{\gamma e^{in\theta
}-\gamma}{\sqrt{2}}\rangle|\frac{\gamma e^{in\theta}+\gamma}{\sqrt{2}}\rangle.
\end{align}
If the amplitude $|\gamma|$ of the beams is large enough, the Poisson peaks of
the states $|\frac{\gamma e^{in\theta}-\gamma}{\sqrt{2}}\rangle$ in the above
output can be mutually separated; see \cite{Lin2}. Then, for the different
number $n$ occurring with the probabilities $e^{-2|\alpha\sin\theta|^{2}}
\frac{|\sqrt{2}\alpha\sin\theta|^{2n}}{n!}$, a detector without the capability
of resolving photon numbers could respond distinguishably to the measured beam
probabilistically in the states $|\zeta_{n}\rangle=|\frac{\gamma e^{in\theta
}-\gamma}{\sqrt{2}}\rangle$, realizing the photon number resolving detection
in an indirect way. The mutually distinct readings of the detector could be a
monotonic function of the expectation values $\langle\zeta_{n}|\hat{\Pi}%
|\zeta_{n}\rangle$, where $\hat{\Pi}=\sum_{m} \{1-(1-\eta)^{m}\}|m\rangle
\langle m|$ is the the positive-operator-value measure (POVM) element
describing the action of a photon number non-resolving detector with the
efficiency $\eta$. Note that, due to the finite range of photon numbers $n$
for the Poisson distribution of $|\sqrt{2}\alpha\sin\theta\rangle$, the
readings of the detector are virtually finite too. The error probability in
one operation of such QND module is \cite{Lin2, Lin1}
\begin{equation}
P_{E}\sim exp\{-2(1-e^{-\frac{1}{2}\eta\gamma^{2}\theta^{2}})\alpha^{2}%
\sin^{2}\theta\}. \label{probability}%
\end{equation}
Even if $\theta\ll1$ due to the weak cross-Kerr nonlinearity, the operation
can be effectively deterministic given $\alpha\sin\theta\gg1$ and
$\gamma\theta\gg1$.

Going back to the operation of our entangler, we will finally apply a Hadamard
operation on the spider photon $a$ to yield the output state
\begin{align}
|\Psi\rangle_{out}  &  = \left(  \left\vert \psi_{1}\right\rangle \left\vert
0\right\rangle \left\vert 0\right\rangle \left\vert +\right\rangle +\left\vert
\psi_{1}\right\rangle \left\vert 0\right\rangle \left\vert 1\right\rangle
\left\vert -\right\rangle +\left\vert \psi_{2}\right\rangle \left\vert
1\right\rangle \left\vert 0\right\rangle \left\vert +\right\rangle \right.
\nonumber\\
&  \left.  -\left\vert \psi_{2}\right\rangle \left\vert 1\right\rangle
\left\vert 1\right\rangle \left\vert -\right\rangle \right)  _{pra}.
\label{out}%
\end{align}
If the spider photon $a$ were projected out by a measurement on $|0\rangle
_{a}$ or $|1\rangle_{a}$ basis, the completed operation by the entangler would
be $CZ_{p,r}$ (the output due to the projection on $|1\rangle_{a}$ should be
modified with a single-qubit operation $\sigma_{z}$ on the target photon $r$),
which connects the entanglement bond between photon $p$ and $r$.

\begin{figure}[ptb]
\includegraphics[width=7.7cm]{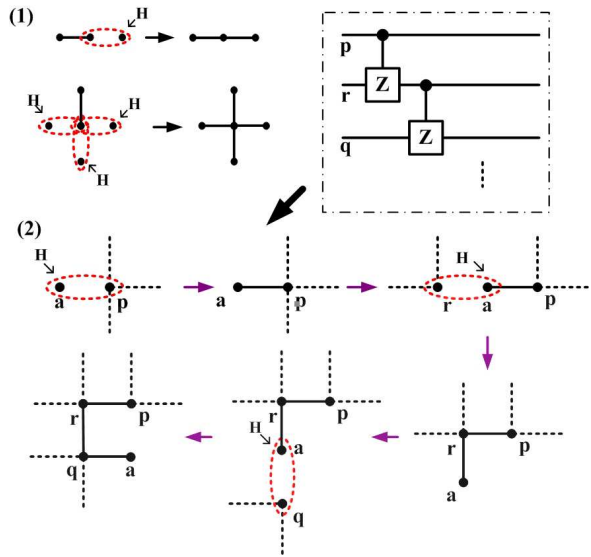}\caption{(color online) (1)
Preparation of a linear and a star-shaped graph state. The red circle denotes
an entangler operation, and H a Hadamard operation. The initial building
blocks for the graph states are single photon qubits, as well as Bell pairs
which can be prepared by the entangler in Fig.1 without using ancilla. (2)
Realization of a 2-cascade CZ operation. The entangler first entangles photon
$a$ and photon $p$ by putting $p$ at the position of $r$ in Fig. 1. Then,
photon $a$ assists the connection of the entanglement bonds between photon $p$
and $r$, $r$ and $q$ in succession. It acts like a spider weaving these
photons together into a string. A general cascade CZ operation like that in
the dash-dotted line can be implemented in this fashion.}%
\end{figure}

In the input state of Eq. (\ref{in}), photon $a$ is specifically correlated to
photon $p$ such that $|+\rangle_{a}$ ($|-\rangle_{a}$) is in the same terms
with $|0\rangle_{p}$ ($|1\rangle_{p}$). After the operation described above,
such correlation is transfered to between photon $a$ and $r$; see Eq.
(\ref{out}). One could use the output state as the input in the form of Eq.
(\ref{in}) to entangle the next photon to photon $r$, and so on. The
operations of entangling a number of independent photons to a string can be
therefore assisted by the same spider photon $a$, as the specific entanglement
with the spider photon is transferred from photon to photon; see part (2) of
Fig. 2 for a basic $2$-cascade entangler operation. The spider photon is the
only ancilla for implementing a cascade entangler operation, and it will not
be destroyed before we complete the whole operation.

Obviously, the starting piece or the Bell pair between the first photon
$p_{1}$ and the spider photon $a$ (here we have no other photons in $|\psi
_{1}\rangle, |\psi_{2}\rangle$ of Eq. (\ref{in})) can be prepared by the same
entangler as well. Now photon $p_{1}$ should take the position of photon $r$
in Fig. 1. The coupling of the cross-Kerr nonlinearities in the same pattern,
as well as the detection by the QND module, transforms the state
$|+\rangle_{p_{1}}|+\rangle_{a}$ to either $|0\rangle_{p_{1}}|0\rangle
_{a}+|1\rangle_{p_{1}}|1\rangle_{a}$ or $|0\rangle_{p_{1}}|1\rangle
_{a}+|1\rangle_{p_{1}}|0\rangle_{a}$, which can be converted to $|0\rangle
_{p_{1}}|+\rangle_{a}+|1\rangle_{p_{1}}|-\rangle_{a}$ by local operations.
Compared with the strategy in \cite{Lin3}, all operations here can be
performed by the same entangler in Fig. 1.

Moreover, extra entanglement links between qubits in an already connected
piece can be built up by the entangler. One example of this case is that
photon $p$ and $r$ are entangled to a third photon so that their total state
is not separable. The entangler will operate in the same way except for
processing a more general input like
\begin{align}
|\Psi\rangle_{in}  &  \left.  = \left\vert \psi_{00}\right\rangle \left\vert
0\right\rangle _{p} \left\vert 0\right\rangle _{r} \left\vert +\right\rangle
_{a} +\left\vert \psi_{01}\right\rangle \left\vert 0\right\rangle _{p}
\left\vert 1\right\rangle _{r} \left\vert +\right\rangle _{a} \right.
\nonumber\\
&  \left.  +\left\vert \psi_{10}\right\rangle \left\vert 1\right\rangle _{p}
\left\vert 0\right\rangle _{r} \left\vert -\right\rangle _{a} +\left\vert
\psi_{11}\right\rangle \left\vert 1\right\rangle _{p} \left\vert
1\right\rangle _{r} \left\vert -\right\rangle _{a} . \right.  \label{in-2}%
\end{align}
This implies that spider photons can make circles by going through a same
qubit in graph states for more than one time.

There is also a linear optical version for the cascade entangler; see Fig. 3.
By the photon-photon interference via the polarization beam splitters and the
coincident photon detection, the linear optical circuit effectively realizes a
CZ operation as the nonlinear one in Fig. 1. Such linear optical cascade
entangler only works with a success probability $1/2^{n}$ in generating a
string of $n$ photonic qubits, but it could be used for experimental
demonstration of this type of entangler.

\begin{figure}[ptb]
\includegraphics[width=5.7cm]{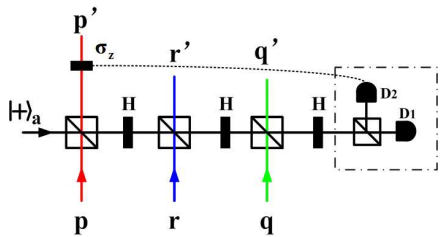}\caption{(color online) Linear optical
realization of the cascade CZ operation. Via the polarization beam splitters,
the spider photon in the state $\left\vert +\right\rangle $ interferes with
the photons p, r, q in succession. The PBS act as entanglers or parity gate,
thus realizing a 2-cascade CZ operation associated with Hadamard operations
performed on the ancilla between two PBS and the coincidence measurement.}%
\end{figure}

\section{Generation of graph states}

In the generation of graph states, we stipulate that the spider photons should
move along the paths that do not retrace an already connected link; otherwise
it will destroy the bonds by the operation of entangler. This requirement is
the same as that for walking through the edges of a graph once and only once;
c.f. the problem of Seven Bridges of K{\"{o}}nigsberg, a notable mathematics
problem solved by L. Euler.

The generation strategy we have presented can be viewed as weaving up a graph
network by spider photons following the above rule. One could have an optimal
decomposition of a graph state, such that the number of parallel or separate
operations for generating the state is minimized by letting one spider photon
go through as many vertexes as possible. This spider photon could pass through
a qubit in graph states for any number of times as long as it obeys the rule
of not stepping on the already connected links, as a cascade entangler is able
to process the most general input in Eq. (\ref{in-2}). In what follows we will
give a few examples of generating graph states by the weaving strategy.

The first example is a square-shaped 2D graph state. Without loss of
generality, we demonstrate the preparation of a $5\times5$ one in Fig. 4. The
initial building blocks, the short chains of two or three qubits, can be
prepared by the same cascade entangler beforehand or generated simultaneously
with different entanglers, while the unlinked qubits will be connected to a
single string from one of the initially linked. During the operation, the
unlinked qubits are woven by a spider photon (not shown explicitly in the
figure) in turn to the string, as it makes the indicated paths going through
the qubits inside the square for more than one time. This string covers all
entanglement links except for those initially connected.

\begin{figure}[ptb]
\includegraphics[width=5.7cm]{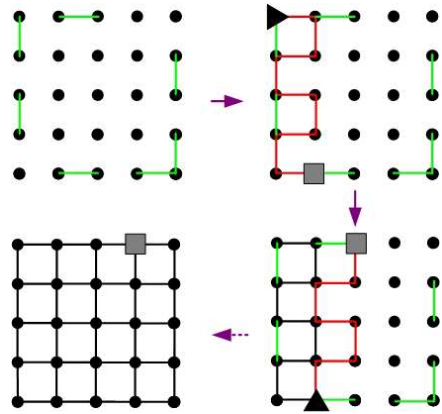}\caption{(color online) Generation of
$5\times5$ graph state. Some linear graph states (denoted by green lines) are
prepared at the beginning as the building blocks. Then, one ancilla photon
(not shown explicitly), which is entangled to one of the linear graph states,
is used as the spider photon to connect the entanglement bonds. In each step,
its starting point is marked by a triangle and its ending point marked by a
square. The paths of the spider photon are denoted by red lines.}%
\end{figure}

\begin{figure}[ptb]
\includegraphics[width=5.7cm]{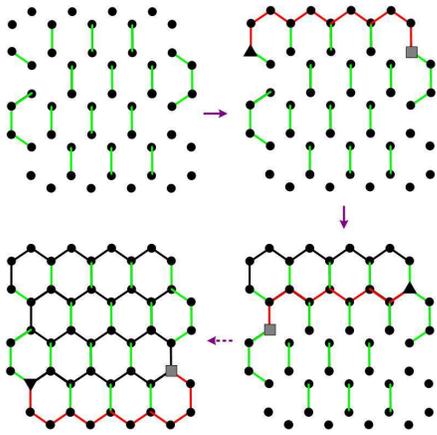}\caption{(color online) Generation
of $4\times4$ alveolate graph state. Some linear graph states are used as the
building blocks. In each step, the starting point of the spider photon is
marked by a triangle and its ending point marked by a square.}%
\end{figure}

The generation of alveolate graph state \cite{dc-c1, alveolate}, a special
kind of 2D graph state, is similar. Following the rule for the spider photon,
a general alveolate graph state can be generated as it walks through the paths
denoted by red lines in Fig. 5.

\begin{figure}[t]
\includegraphics[width=5.7cm]{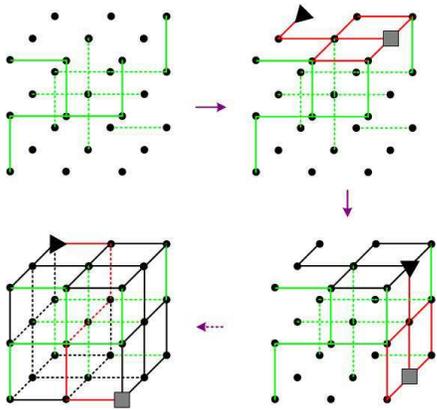}\caption{(color online) Generation of
$3\times3\times3$ graph state. The building blacks used here are the
star-shaped and linear graph states prepared as in Fig. (2.1), as well as the
individual photonic qubits. In each step, the starting point of the spider
photon is marked by a triangle and its ending point marked by a square.}%
\end{figure}

\begin{figure}[b]
\includegraphics[width=0.5\textwidth]{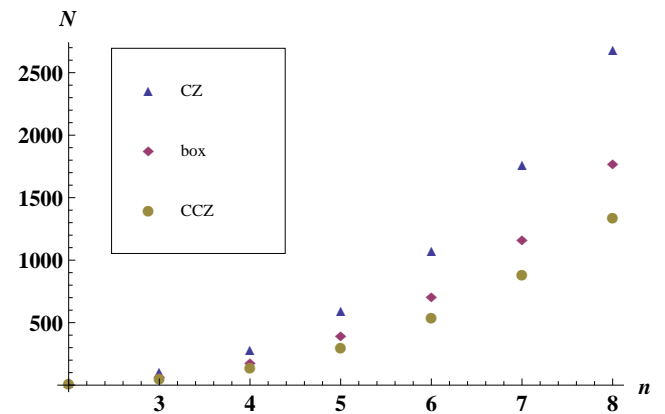}\caption{(color online)
Comparison of the numbers of the individual entangler operations in generating
an $n\times n \times n$ cubic graph state. The number by the cascade entangler
(CCZ) in this paper is $N=3(n-1)n^{2}+1$. By the block by block (box) strategy
in \cite{Lin3}, the corresponding number is $n=4(n-1)n^{2}-2n+1$, if $n$ is
even; and $N=4(n-1)n^{2}-n+1$, if $n$ is odd. The corresponding number by the
direct CZ operations on each entanglement bond is $N=6(n-1)n^{2}$.}%
\end{figure}

The advantage of the weaving strategy is more obvious in 3D graph state
generation. 3D graph states are proposed for the fault-tolerant one-way
quantum computing \cite{3D}. The three dimension of a graph state means the
cubically increasing entanglement bonds, which demand much more operations
than in preparing 2D graph states. Here we use the example of $3\times
3\times3$ graph state in Fig. 6 to illustrate the generation of 3D graph state
by cascade CZ operations. The building blocks are two star-shaped graph states
and one linear graph state prepared as in Fig. (2.1), as well as a Bell pair
and the rest unconnected single-photon qubits. The star-shaped graph states
and Bell pair are denoted with the green bonds in Fig. 6. All these pieces can
be generated by the simultaneous operations of more entanglers, or by the same
entangler beforehand or afterward. The passage of the spider photon is shown
with the red lines in each step of Fig. 6. The spider photon keeps traveling
side by side on the cubic until it connects all entanglement links other than
the initially connected.

Only counting the number of the individual entangling operations in generating
an $n\times n\times n$ graph state, this weaving strategy improves on other
approaches by requiring less operations; see Fig. 7. Such improvement is due
to the fact that the passage of a spider photon can be on different 2D planes
of a cubic. Since the operations in generating each string $E_{i}$ can be
bundled together by a cascade entangler, the actually number of separate
operations in preparing an $n\times n\times n$ cubic graph state is in the
order of that of the linked initial building blocks in Fig. 6, whose quantity
is in the order of $n$. By our weaving strategy, therefore, the quantity of
the necessary resource for preparing a 3D graph state only grows with the size
number $n$ rather than its link number $n^{3}$.

\section{Discussion and conclusion}

The core element for our cascade entangler is a proper weak cross Kerr
nonlinearity for entangling operations and in QND modules. The recent
experimental progress on photonic XPM can be found in, e. g., \cite{chen,
fiber, lo, ex4, ex5}. More progress in the research of such photonic
nonlinearity is expected in the near future.

Most of theoretical studies thus far (including the present one) adopt the
simplified single mode treatment for XPM between photons and coherent states.
This picture is criticized in \cite{shap2}, which considers the quantum noise
effect due to the non-instantaneous response of nonlinear media. However, in
the regime of very small conditional phase $\theta$, a multi-mode XPM from the
instantaneous photonic interactions works almost in the same way as that of
single mode approximation \cite{multi, qft}. Given a small conditional phase
$\theta\ll1$, our deterministic entangler should work in the regime of
$\alpha\sin\theta\gg1$. Such compatibility between a small conditional phase
and a large average photon numbers of the qubus beams in the realistic
multi-mode XPM is clarified in \cite{xpm}, where we adopt a continuous-mode
interacting quantum field model to show the validity of the single mode
approximation in this regime.

We have illustrated how to apply cascade entangler operations to entangle
independent single photons into arbitrary graph state. This approach is highly
efficient and suitable to preparing the graph states of large number of
qubits. It could make photon a suitable qubit for the realistic one-way
quantum computing.

\begin{acknowledgments}
The authors thank Dr. Jianming Cai and Ru-Bing Yang for helpful suggestions.
Q. L. was funded by National Natural Science Foundation of China (Grant
No.11005040), the Natural Science Foundation of Fujian Province of China
(Grant No.2010J05008) and the Fundamental Research Funds for Huaqiao
University (Grant No. JB-SJ1007). B. H. acknowledges the support by Alberta Innovates.
\end{acknowledgments}

\end{document}